\documentclass[12pt,preprint]{aastex}

\usepackage{rotating}

\shorttitle{X-Ray Spectral Evolution of SN~2005ip} 
\shortauthors{Katsuda et al.}

\begin{document}

\title{SN~2005ip: A Luminous Type~IIn Supernova Emerging from a Dense Circumstellar Medium as Revealed by X-Ray Observations}

\author{Satoru Katsuda\altaffilmark{1}, Keiichi Maeda\altaffilmark{2, 3}, 
Takaya Nozawa\altaffilmark{3}, David Pooley\altaffilmark{4, 5}, 
\& Stefan Immler\altaffilmark{6, 7, 8}
}

\altaffiltext{1}{RIKEN (The Institute of Physical and Chemical Research) Nishina Center, 2-1 Hirosawa, Wako, Saitama 351-0198, Japan}

\altaffiltext{2}{Department of Astronomy, Kyoto University, Kitashirakawa-Oiwake-cho, Sakyo-ku, Kyoto 606-8502, Japan}

\altaffiltext{3}{Kavli Institute for the Physics and Mathematics of the Universe (WPI), University of Tokyo, 5-1-5 Kashiwanoha, Kashiwa, Chiba 277-8583, Japan}

\altaffiltext{4}{Department of Physics, Sam Houston State University, Huntsville, Texas 77341-2267, USA}

\altaffiltext{5}{Eureka Scientific, Inc., 2452 Delmer Street, Suite 100, Oakland, CA 94602, USA}

\altaffiltext{6}{Astrophysics Science Division, NASA Goddard Space Flight Center, Greenbelt, MD 2077, USA}

\altaffiltext{7}{Center for Research and Exploration in Space Science and Technology, NASA Goddard Space Flight Center, Greenbelt, MD 20771, USA}

\altaffiltext{8}{Department of Astronomy, University of Maryland, College Park, MD 20742, USA}







\begin{abstract}
We report on X-ray spectral evolution of the nearby Type~IIn supernova (SN) 2005ip, based on {\it Chandra} and {\it Swift} observations covering from $\sim$1 to 6 years after the explosion.  X-ray spectra in all epochs are well fitted by a thermal emission model with $kT\gtrsim7$\,keV.  The somewhat high temperature suggests that the X-ray emission mainly arises from the circumstellar medium heated by the forward shock.  We find that the spectra taken 2--3 years since the explosion are heavily absorbed ($N_{\rm H}\sim5\times10^{22}$\,cm$^{-2}$), but the absorption gradually decreases to the level of the Galactic absorption ($N_{\rm H}\sim4\times10^{20}$\,cm$^{-2}$) at the final epoch.  This indicates that the SN went off in a dense circumstellar medium and that the forward shock has overtaken it.  The intrinsic X-ray luminosity stays constant until the final epoch when it drops by a factor of $\sim$2.  The intrinsic 0.2--10\,keV luminosity during the plateau phase is measured to be $\sim1.5\times10^{41}$\,erg\,s$^{-1}$, ranking SN~2005ip as one of the brightest X-ray SNe.  Based on the column density, we derive a lower-limit of a mass-loss rate to be $\dot M\sim1.5\times10^{-2}$\,($V_w/100$\,km\,s$^{-1}$)\,$M_\odot$\,yr$^{-1}$, which roughly agrees with that inferred from the X-ray luminosity, $\dot M\sim2\times10^{-2}$\,($V_w/100$\,km\,s$^{-1}$)\,$M_\odot$\,yr$^{-1}$, where $V_{w}$ is the circumstellar wind speed.  Such a high mass-loss rate suggests that the progenitor star had eruptive mass ejections like a luminous blue variable star.  The total mass ejected in the eruptive period is estimated to be $\sim$15\,$M_\odot$, indicating that the progenitor mass is $\gtrsim25$\,$M_\odot$.
\end{abstract}


\section{Introduction}

Type~IIn supernovae (SNe~IIn) are a class of supernovae (SNe) that show prominent narrow H emission lines in their optical spectra \citep{Schlegel1990}.  There is a general consensus that these SNe occur in dense circumstellar media (CSM) created by pre-SN mass losses, and that the narrow H lines are produced by photoionization of the dense winds irradiated by X-rays from the region behind the forward shock.  However, the detailed nature of SNe~IIn is still unclear, because well-studied objects are sparse \citep[][for recent reviews]{Taddia2013,Kiewe2012}.  

In particular, their progenitors have not yet been established.  One of the most important quantities to constrain the progenitor is a mass-loss rate, which varies from $\sim$10$^{-6}$--10$^{-5}\,M_\odot$\,yr$^{-1}$ for red supergiants (RSG), $\sim$10$^{-5}$--10$^{-4}\,M_\odot$\,yr$^{-1}$ for stripped Wolf-Rayet stars, and to $\sim$10$^{-4}$--10\,$M_\odot$\,yr$^{-1}$ for luminous blue variable (LBV) stars \citep{Kiewe2012}.  We can infer mass-loss rates of progenitor stars by measuring H$\alpha$ and X-rays from SNe, since they arise due to interactions between the CSM and ejecta.  So far, mass-loss rates have been inferred for several SNe~IIn.  The values scatter within 10$^{-4}$--10\,$M_\odot$\,yr$^{-1}$, consistent with massive eruptions seen for LBV stars \citep{Kiewe2012,Taddia2013}.  Indeed, possible detections of LBV-like progenitors in pre-SN images were reported for SNe~2005gl and 2009ip \citep{Gal-Yam2009,Mauerhan2013,Pastorello2013}.  

SN~2005ip, discovered on 2005 November 5.163 in the nearby galaxy NGC~2906 \citep{Boles2005}, was initially designated a normal Type~II SN, based on an optical spectrum taken 1 day after discovery \citep{Modjaz2005}.  But later, \citet{Smith2009b} found clear narrow H$\alpha$ emission even on day~1 as well as an unusually rich forest of narrow coronal lines at later times, and categorized SN~2005ip as SNe~IIn.  \citet{Fox2009} reported strong IR emission over more than 900 days post discovery.  Subsequent studies on IR photometric and spectroscopic observations showed two dust components, one presumably formed in the ejecta or cool dense shell and the other pre-existing dust in the surrounding CSM \citep{Fox2010}.  X-ray emission from SN~2005ip was detected on 2007 November with {\it Swift} \citep{Immler2007b}.  For an assumed foreground column density of $N_{\rm H}$ = 3.7$\times10^{20}$\,cm$^{-2}$ and a thermal spectrum with a plasma temperature of $kT = $ 1\,keV, an absorption-corrected (or intrinsic) X-ray luminosity was calculated to be 1.6$\pm0.3\times10^{40}$\,erg\,s$^{-1}$ \citep{Immler2007b}.  

Previous optical and IR observations have led to different estimates of mass-loss rates for the progenitor of SN~2005ip.  From the H$\alpha$ intensity together with the CS wind speed ($V_w = 100$\,km\,s$^{-1}$), \citet{Smith2009b} inferred that $\dot M =$ (2--4)$\times10^{-4}$\,$M_\odot$\,yr$^{-1}$ and argued that the progenitor is a RSG star having eruptive mass ejections like VY Canis Majoris \citep{Smith2009a}.  By contrast, a large amount of warm dust contained in the CS shell led \citet{Fox2010} to suggest a very high mass-loss rate of 0.075--0.38\,$M_\odot$\,yr$^{-1}$.  Such a mass-loss rate exclusively favors a LBV-like progenitor rather than a RSG.  Most recently, \citet{Moriya2013} estimated that $\dot M =$ (1.2--1.4)$\times10^{-3}$\,$M_\odot$\,yr$^{-1}$, based on analytical modeling of the bolometric light curve.  This result further supports a LBV-like progenitor.

In this paper, we report on archival X-ray monitoring observations of SN~2005ip from $\sim$1\,yr to $\sim$6\,yr post discovery, acquired by {\it Swift} and {\it Chandra}.  X-ray emission has been detected in all the observations, exhibiting clear spectral softening with time.  The spectral evolution can be best interpreted by a significant decline of absorption due to the external CSM.  The X-ray spectra enable us to measure a correct X-ray luminosty as well as an absorbing column density, allowing for a new estimate of the mass-loss rate, independent of the other previous ones.  Throughout this paper, we use a distance to SN~2005ip of 35\,Mpc \citep{Stritzinger2012}, and adopt 2005 October 27.2 (9 days prior to the discovery) as the time of explosion \citep{Smith2009b}.


\section{Observations and Analysis}

We analyze several archival {\it Swift} and {\it Chandra} data for SN~2005ip using the HEAsoft\footnote{http://heasarc.gsfc.nasa.gov/docs/software/lheasoft/} and CIAO\footnote{http://asc.harvard.edu/ciao/} packages.  Detailed information of these observations is summarized in Table~\ref{tab:obs}.  We combine the latest five {\it Swift} observations taken in 2012 March into one data file owing to the poor photon statistics for individual data sets.  As a result, we have five observation epochs, i.e., 2007 February ({\it Swift}), 2008 June ({\it Chandra}), 2008 November ({\it Swift}), 2009 November ({\it Swift}), and 2012 March ({\it Swift}).  

\begin{figure}[htbt]
\begin{center}
\includegraphics[angle=0,scale=0.65]{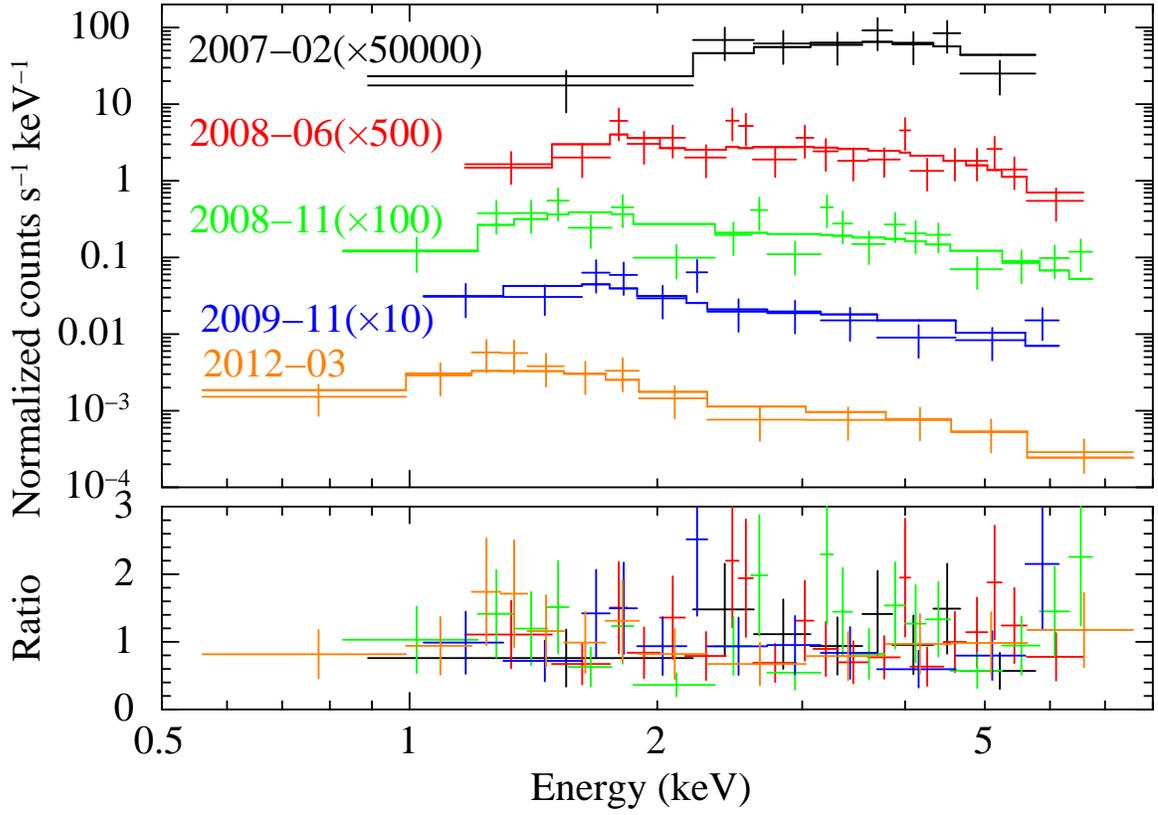}\hspace{1cm}
\caption{X-ray spectra along with the best-fit models ({\tt TBabs}$\times${\tt apec}).  All the data were obtained with the {\it Swift} XRT except for the one in 2008 June which was taken with the {\it Chandra} ACIS-S.  The lower panel shows the ratios of the data to the best-fit model.
} 
\label{fig:spec}
\end{center}
\end{figure}

We extract X-ray spectra from a circular region with a radius of 30$^{\prime\prime}$ (3$^{\prime\prime}$) for {\it Swift} ({\it Chandra}) observations, and subtract background from the surrounding annular region.  The background levels to the source are quite small: $\sim$0.3\% ({\it Chandra}) and $\sim$3\% ({\it Swift}).  Each spectrum is grouped into bins, so that each bin contains at least 5 counts.  We have checked that even if we set the number of the minimum grouping count to be unity, we obtain the same fit results shown below.  The background-subtracted spectra are shown in Fig.~\ref{fig:spec}.  Clearly, the spectra are softening with time, as is evident in hardness ratios summarized in Table~\ref{tab:obs}.

\begin{figure}
\begin{center}
\includegraphics[angle=0,scale=0.45]{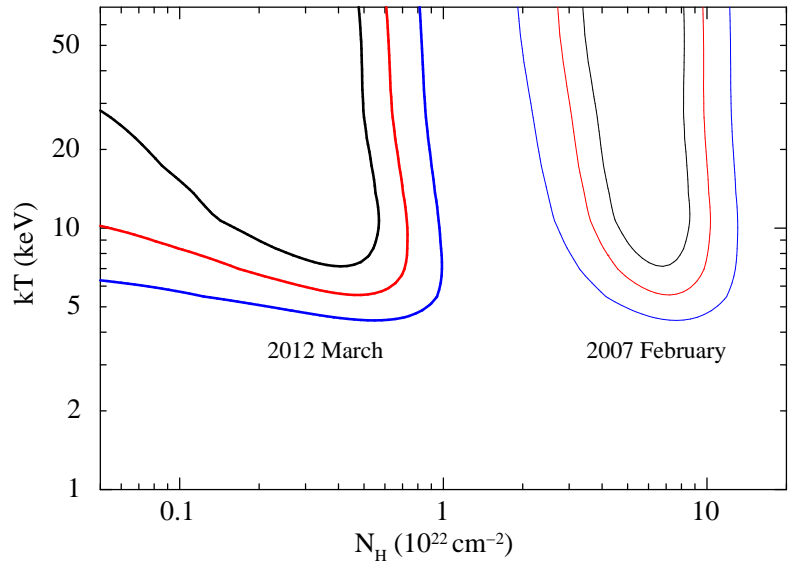}\hspace{1cm}
\caption{Confidence contours of $kT$ vs.\ $N_{\rm H}$, for which black, red, and blue represent confidence levels of $\Delta C = 2.3$, 4.6, and 9.2 (corresponding to 68\%, 90\%, and 99\% for two interesting parameters).  Thin and thick lines correspond to the results taken in 2007 February and 2012 March, respectively.  
} 
\label{fig:conf_cont}
\end{center}
\end{figure}

For spectral modeling, we apply an absorbed \citep[{\tt TBabs}:][]{Wilms2000}, thermal emission model in ionization equilibrium, i.e., the {\tt apec} model \citep{Smith2001} in XSPEC \citep{Arnaud1996}.  We allow the absorbing H column density ($N_{\rm H}$), the temperature ($kT$), and normalizations to be free, while we treat the temperature to be common among all the five spectra in order to better constrain the parameters.  Elemental abundances in both {\tt TBabs} and {\tt apec} models are set to the solar values \citep{Lodders2003}, which are consistent with the metallicity of the host galaxy NGC~2906 \citep{Stritzinger2012}.  To find the best-fit model, we use maximum likelihood statistics for a Poisson distribution, the so-called c-statistics \citep{Cash1979}, which minimizes the $C$ value defined as $C = -2\sum_i(M_i - D_i + D_i ({\rm ln} D_i - {\rm ln} M_i))$ with $M_i$ and $D_i$ being the model-predicted and observed counts in each spectral bin $i$, respectively.  This method is suitable for low-counts spectra that do not allow for the $\chi^2$-test.  Errors on the parameters can be estimated similarly to the $\chi^2$ method, based on the fact that $\Delta C = C - C_{\rm min}$ can be used similarly to $\Delta \chi^2$ \citep{Cash1979}, where $C_{\rm min}$ is the minimum $C$ value obtained at the best-fit.  As can be seen in Fig.~\ref{fig:spec}, the best-fit models represent the data well.  The best-fit parameters are summarized in Table~\ref{tab:param1} (No.1).  We find a clear temporal decrease in $N_{\rm H}$.  This result is fairly robust from a statistical point of view, as is evident from confidence contours of $kT$ vs.\ $N_{\rm H}$ in Fig.~\ref{fig:conf_cont}, where thick and thin lines correspond to the first and the final epochs, respectively.  Judging from the high temperature ($\gtrsim$6.7\,keV), most of the X-ray emission likely arises from a hot plasma in the CSM heated by the forward shock.  

We also fit the data with the same model in different ways that take account of temperature evolution.  First, we assume that the temperature evolves as $T \propto t^{-0.2}$, since the temperature is proportional to the square of the shock speed which decreases as $t^{-0.1}$ in a self-similar solution \citep[e.g.,][]{Chevalier1982}; the SN radius, $R$, is proportional to $t^{(n-3)/(n-s)}$, where $n$ and $s$ are density profiles in the ejecta and the CSM, respectively ($\rho_{\rm ej} \propto r^{-n}$ and $\rho_{\rm csm} \propto r^{-s}$), so that the shock speed, $dR/dt$, is proportional to $t^{(n-3)/(n-s)-1}$, resulting in $dR/dt \sim t^{-0.1}$ for typical density profiles of core-collapse SNe, $n \sim 10$ and $s \sim 2$ \citep[e.g.,][]{Chevalier2003}.  The fit results listed in Table~\ref{tab:param1} (No.2) are found to be consistent (except for the temperature) with those of the No.1 case.  Second, we take into account effects of temperature nonequilibration between electrons and ions due to slow equilibration through Coulomb heating.  The temperature nonequilibration may be applicable for SN~2005ip, in which X-rays are likely due mainly to an adiabatic forward shock \citep[e.g.,][]{Fransson1996}.  To this end, we follow \citet{Nymark2009} who showed that electron temperature is proportional to $V_{\rm sh}^{-4/3} t^{-2/3}$ for Coulomb heating \citep[Equation 10 in][]{Nymark2009}.  Using the relation $V_{\rm sh} \propto t^{-0.1}$, we obtain $T_{\rm e} \propto t^{-0.53}$.  By fitting the five spectra with this temperature evolution taken account, we obtain the results shown in Table~\ref{tab:param1} (No.3), which are consistent (except for the tempetarue) with the other cases.  Finally, we allow $kT$ to vary freely among the five spectra, while we let $N_{\rm H}$ be a common free parameter.  This strategy fails to fit the data ($C$ value of 93.1; the fit results are presented in Table~\ref{tab:param1} No.4), indicating that the spectral softening is not simply due to cooling.  Therefore, we are confident that the spectral softening is mainly due to evolution in the absorption.  

While nondetection of X-ray emission from reverse-shocked ejecta is not so peculiar for SNe~IIn \citep[cf.][]{Chandra2012a,Chandra2012b}, for completeness we also apply an absorbed two-components {\tt apec} model, in which the two components are supposed to represent the reverse-shocked ejecta and the forward-shocked CSM \citep[e.g.,][but see also \citet{Nymark2006,Nymark2009} who presented a more detailed modeling especially focusing on X-ray emission from the ejecta by coupling a spectral code to a hydrodynamical code]{Fransson1996}.  Since it is difficult to determine a number of spectral-fit parameters with the relatively poor photon statistics, we initially link temperatures and normalizations of the low-temperature (ejecta) component among the five spectra, according to a theoretical expectation by \citet{Fransson1996}.  The best-fit parameters are summarized in Table~\ref{tab:param2} (No.1).  The two-component model gives a slightly better fit than the one-component model: $\Delta C \sim -$4.5 for two extra free parameters (i.e., $kT$ and normalization of the ejecta component).  The slight improvement in the fit is not strongly significant, yielding a confidence level of 92\% based on the $F$-test; we assume that the $C$ values can be treated as $\chi^2$ values and performed $F$-test using the fit-statistics listed in Table~\ref{tab:param2} (No.1).  It should be noted that, as can be seen from the best-fit parameters listed in Tables~\ref{tab:param1} and \ref{tab:param2}, the basic best-fit parameters, i.e., $N_{\rm H}$, $kT$, and normalization of the high-temperature component, derived by the one- and two-component models roughly agree with each other.  

Next, to see if the spectral softening can be explained by the increasing low-temperature component rather than the $N_{\rm H}$ evolution, we fit the data with the two-component model in a different fitting strategy, in which we link the $N_{\rm H}$ parameter among the five epochs while we let the normalization of the low-temperature component vary freely for individual epochs.  The best-fit parameters are listed in Table~\ref{tab:param2} (No.2), from which we find that the model gives a reasonable but slightly worse fit than previous fittings.  Moreover, the intrinsic X-ray luminosity for the low-temperature component is derived to be $\sim$10$^{43}$\,ergs\,s$^{-1}$ at the final epoch ($\sim$6 yr after explosion).  This may be unrealistically bright especially for reverse-shock emission; it is two orders of magnitude brighter than ever recorded at this late phase \citep{Dwarkadas2012}.  The intrinsic X-ray luminosity for the cooler component would be even greater than the value we obtained.  This is because there is almost always a radiative cooling shell between a reverse shock and a forward shock, and the cooling shell will provide an extra absorption of the soft X-rays coming from the reverse shock.  We have also checked that temperature evolution of $T \propto t^{-0.2}$ for both of the two components (as is expected in a reasonable self-similar solution described above) does not affect the fit result (see the results No.3 in Table~\ref{tab:param2}).  Accordingly, in the following discussion, we basically adopt the results from the one-component model in Table~\ref{tab:param1}.

\section{Discussion}

The archival X-ray data of SN~2005ip revealed clear spectral softening with time.  Our spectral analysis showed that the spectral softening is best explained by declining evolution of the absorbing column density; $N_{\mathrm H} \sim 5\times10^{22}$\,cm$^{-2}$ at the first epoch ($\sim$1\,yr after explosion) gradually goes down to be consistent with the Galactic absorption, $N_{\rm H} \sim 4\times10^{20}$\,cm$^{-2}$, at the final epoch ($\sim$6\,yr after explosion).  On the other hand, we find that the intrinsic X-ray luminosity stays constant until the final epoch when it drops by a factor of $\sim$2 (see Table~\ref{tab:param1}).  These results suggest that the SN was initially embedded in a dense CSM, but the forward shock has penetrated the dense CSM region at a certain time between 2009 November (fourth epoch) and 2012 March (final epoch). 

SN~2005ip is the second example in which we see possible evolution in absorption by the external CSM after SN~2010jl \citep{Chandra2012a}.  By contrast, there are several SNe that do not show significant $N_{\rm H}$ evolution.  These include SNe 2006jd \citep{Chandra2012b}, 1987A \citep{Dewey2012}, 1998S \citep{Pooley2002}, and 1999em \citep{Pooley2002}.  SN~1993J would be intermediate between the two types, i.e., evolving absorption and constant absorption; \citet{Chandra2009} noted possible $N_{\rm H}$ evolution that may be caused by a cool shell between the forward shock and the reverse shock.  A simple speculation to explain the difference between the two types is that the CSM is formed as a disk-like shell due to equatorial mass ejections.  For this geometry, CSM absorption strongly depends on the inclination angle of the disk in the sense that strong absorption and its declining evolution is expected for an edge-on view (which would be responsible for SNe 2005ip and 2010jl), while little absorption for a face-on view (for others).  An alternative scenario is that the class of SNe without evolution in absorption has somewhat peculiar CSM geometries (while the SNe class with strongly evolving absorption has either a relatively spherically symmetric CSM or an edge-on disk).  For example, \citet{Chandra2012b} proposed for SN~2006jd that a low-column-density region is present in one direction (i.e., a cylindrical shape toward the observer) while strong interactions are taking place over much of the rest of the solid angle.  In order to constrain the CSM structures, it is essential to increase SNe samples that allow for absorption measurements with {\it Chandra}, {\it XMM-Newton}, and {\it Swift}.  

\begin{figure}
\begin{center}
\includegraphics[angle=0,scale=0.45]{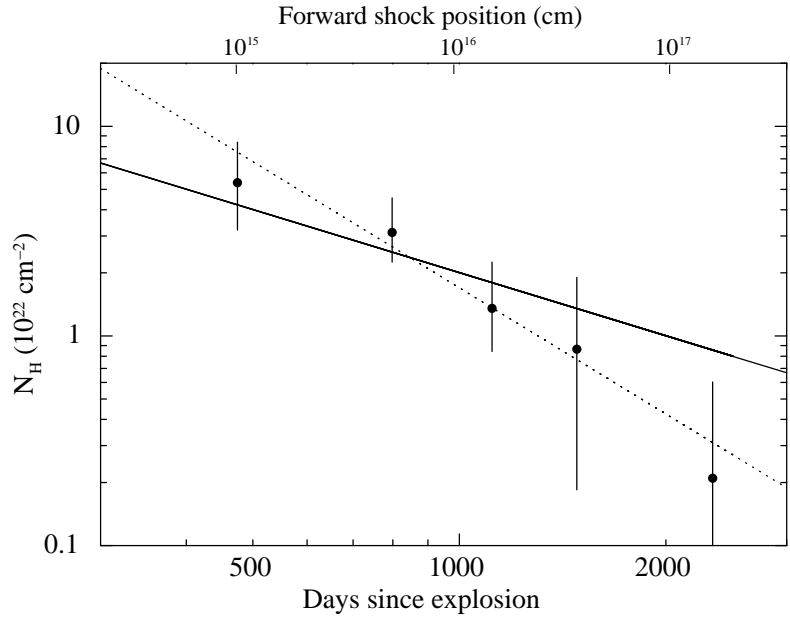}\hspace{1cm}
\caption{$N_{\rm H}$ as a function of time after explosion along with two model curves for either $\dot M = 1.5\times10^{-2} V_{w2} M_\odot$\,yr$^{-1}$ and $s = 2$ (solid) or $\dot M = 4 V_{w2} M_\odot$\,yr$^{-1}$ and $s = 3$ (dotted).  The upper label shows an approximate forward shock position, based on the shock speed of $V_{\rm sh} = 14000$\,($t$/1\,yr)$^{-0.1}$\,km\,s$^{-1}$ \citep[cf.][]{Stritzinger2012}.
}
\label{fig:NH}
\end{center}
\end{figure}

We estimate mass-loss rates of the progenitor star of SN~2005ip in two different ways.  One is based on the column density which is a robust measure in X-ray spectroscopy.  We compare the measured column density with a theoretical calculation, i.e., Equation (4.1) in \citet{Fransson1996}.  As shown in Fig.~\ref{fig:NH}, the data are well represented by either $\dot M = 1.5\times10^{-2}$\,$M_\odot$\,yr$^{-1}$ and $s = 2$ (solid) or $\dot M = 4\,M_\odot$\,yr$^{-1}$ and $s = 3$ (dotted) with reasonable wind and shock speeds \citep[$V_w = $100\,km\,s$^{-1}$ and $V_{\rm sh} = 14000$\,km\,s$^{-1}$:][]{Smith2009b}.  Here, we should not put much weight on the last-epoch data, since the forward shock has likely penetrated the dense CSM region (resulting in a change in the $s$ value) between the fourth epoch and the final epoch.  From an astrophysical point of view, the latter case is not realistic; $\dot M = 4\,M_\odot$\,yr$^{-1}$ is too high and $s = 3$ is too steep especially for Type~IIn SNe which are embedded in usually dense environments.  Therefore, we favor the former case with the mass-loss rate of $\sim$0.015\,$M_\odot$\,yr$^{-1}$ and $s = 2$.

The other way is based on the X-ray luminosity, as is often used to estimate wind parameters in conjunction with Equation (3.11) in \citet{Fransson1996}, which describes X-ray emission from the shocked CSM.  The much larger absorption than that ($N_{\rm H} =3.7\times10^{20}$\,cm$^{-2}$) assumed by \citet{Immler2007b} boosts up the intrinsic 0.2--10\,keV luminosity to be $\sim$1.5$\times$10$^{41}$\,erg\,s$^{-1}$, which is by an order of magnitude higher than the original value of 1.6$\pm0.3\times10^{40}$\,erg\,s$^{-1}$ reported by \citet{Immler2007b}.  The revised luminosity places SN~2005ip as one of the brightest X-ray SNe among others being SNe~2001em \citep{Pooley2004}, 2005kd \citep{Immler2007a,Pooley2007,Dwarkadas2012}, 2006jd \citep{Chandra2012b} and 2010jl \citep{Chandra2012a}.  By integrating the equation from 0.2\,keV to 10\,keV, it can be rewritten as 2.77$\times$10$^{36}$\,(1$-s$)$^{-1}$\,($\dot M_{-4}$/$V_{w2}$)$^{2}$\,($V_{\rm sh}/10^4$\,km\,s$^{-1}$)$^{3-2s}$\,($t$/11.57 days)$^{3-2s}$\,erg\,s$^{-1}$ for $T = 6\times10^{8}$ K, where $\dot M_{-4} = (\dot M / 10^{-4} M_\odot$\,yr$^{-1}$) and $V_{w2} = (V_w/100$\,km\,s$^{-1}$).  The intrinsic luminosity during the plateau phase, 1st--4th epochs when we can reasonably assume that $s=1.5$ from $3-2s=0$, gives a mass-loss rate of $\dot M = (2-2.5)\times10^{-2} V_{w2}$\,$M_\odot$\,yr$^{-1}$, which is in agreement with the lower limit of the $N_{\rm H}$-based $\dot M$ estimated above.  
Therefore, our X-ray--based mass-loss rates are significantly higher than those inferred from the H$\alpha$ luminosity \citep[(2--4$)\times10^{-4}$\,$M_\odot$\,yr$^{-1}$:][]{Smith2009b} and the bolometric light curve \citep[(1.2--1.4)$\times10^{-3}$\,$M_\odot$\,yr$^{-1}$:][]{Moriya2013}, but are close to the estimate from IR observations \citep[(7.5--38)$\times10^{-2}$\,$M_\odot$\,yr$^{-1}$:][]{Fox2010}.  Our result supports eruptive mass ejections as seen in LBV stars.  

Interestingly, the evolution in the X-ray luminosity, which is constant for the initial $\sim$4\,yr and drops between the fourth epoch and the final epoch, is similar to that of H$\alpha$ emission \citep{Stritzinger2012}.  This similarity confirms a general picture that late-time H$\alpha$ emission arises by reprocessing of X-ray emission.  The decreasing X-ray and H$\alpha$ luminosities at the final epoch lead us to suggest that the forward shock broke out a dense CSM region (i.e., entered into a steeper density-profile region) at some point between the fourth epoch and the final epoch.  
The distance that the forward shock \citep[$V_{\rm sh} = 14000$\,km\,s$^{-1}$:][]{Stritzinger2012} traveled for $\sim5$ years, a time span for the forward shock to be within the dense CSM region, is calculated to be $\sim$2.2$\times$10$^{17}$\,cm.  This would be the spatial extent of the erupted material.  Dividing the distance by the wind speed \citep[$V_w \sim 100$\,km\,s$^{-1}$:][]{Smith2009b}, we can estimate the duration of the eruption to be $\sim$700\,yr\,($t$/5\,yr) ($V_{\rm sh}$/14000\,km\,s$^{-1}$) ($V_w$/100\,km\,s$^{-1}$)$^{-1}$.  Then, the total mass ejected in this eruptive period is calculated to be $\sim$14\,($t$/700\,yr) ($\dot M$/0.02\,$M_\odot$\,yr$^{-1}$)\,$M_\odot$, indicating that the progenitor is quite massive ($\gtrsim25$\,$M_\odot$).

As mentioned above, the constant evolution in the X-ray luminosity (the plateau phase) requires that $s = 1.5$ in the framework of general SN evolutionary models \citep[e.g.,][]{Fransson1996}.  At a grance, this appears to conflict with the $N_{\rm H}$-based $s$-value ranging from $\sim$2 to $\sim$3.  However, we should note that such SN evolutionary models assume spherically symmetric CSM structures, which may not be valid for SN~2005ip.  For example, the CSM geometry may be a disk-like shell, as discussed earlier in this section.  In this case, the constant X-ray luminosity is expected for a density profile of $s = 2$ instead of $s = 1.5$ \citep{Stritzinger2012}.  On the other hand, evolution in $N_{\rm H}$ would be similar to that expected for a spherically symmetric CSM, if the disk is edge on.  Therefore, a disk-like CSM structure with $s = 2$ would explain evolution in both $N_{\rm H}$ and luminosity simultaneously without invoking exotic conditions.

There still remains the possibility that the CSM structure of SN~2005ip is spherically symmetric.  In this case, evolution in $N_{\rm H}$ and the luminosity leads to discrepant density profiles, which seems to be a real problem.  By noting that absorbing material decreases more rapidly than what is expected from the luminosity evolution, one may think of a possibility that the external cool CSM is fully ionized by strong X-ray emission behind the forward shock.  We thus calculate the ionization parameter, $\xi = L/nr^{2}$.  For the X-ray luminosity of $L = $1.5$\times10^{41}$\,erg\,s$^{-1}$, a lower limit density of $n = 3\times10^{5}$\,cm$^{-3}$ \citep{Smith2009b}, and a radius of $r = 10^{17}$\,cm which is a distance for a shock of $V_{\rm sh} = 10^4$\,km\,s$^{-1}$ to propagate during $\sim$1000 days, we obtain that $\xi \lesssim 50$.  This is not sufficient to fully ionize C, N, and O which are main absorbers of X-rays \citep[cf.][]{Chandra2012b}.  Therefore, photoionization is unable to explain the discrepancy in the derived density profile.  The other possibility is to introduce clumpiness in the CSM wind, in which shocked clumps sustain the high X-ray luminosity \citep{Chugai1994}, while a covering fraction of dense clumps is so small that they do not noticeably affect dynamics of the system and the CSM absorption.  In this model, a flatter distribution of clumps than that of a global CSM density profile can qualitatively explain both the constant intrinsic luminosity and the rapidly declining $N_{\rm H}$.  Given that the presence of small clumps is strongly suggested by the late-time development of intermediate-width component of H$\alpha$ \citep{Smith2009b}, this scenario may be at work.  The clumps will produce significant soft X-ray emission, which is not required in the one-component model, but is seen in the two-component model (cf. Table~\ref{tab:param2}).  Therefore, the clump scenario cannot be excluded at the expense of a remarkably strong X-ray emission from the reverse shock.


\section{Conclusion}

X-ray spectra from SN~2005ip showed strong spectral softening with time, which we interpret to be due to declining in absorption by external CSM.  With the absorption taken into account, SN~2005ip turned out to be one of the brightest SNe~IIn in the X-ray regime.  Both the column density and the high luminosity require a mass-loss rate of at least $\sim1\times10^{-2} V_{w2}$\,$M_\odot$\,yr$^{-1}$.  Such a high mass-loss rate suggests eruptive mass ejections by a LBV-like progenitor.  Based on evolution in the X-ray luminosity and $N_{\rm H}$, we speculate that the CSM structure of SN~2005ip is a disk-like equatorial shell.  Alternatively, a relatively spherical wind with small clumps could be another possibility at the expense of a remarkably bright reverse shock X-ray emission.

\acknowledgments

S.K.\ is supported by the Special Postdoctoral Researchers Program in RIKEN.  The work by K.M.\ and T.N.\ is supported by World Premier International Research Center Initiative (WPI Initiative), MEXT, Japan.  S.K. and K.M.\ are supported by Grant-in-Aid for Scientific Research for Young Scientists (25800119, 23740141).  T.N.\ has been supported by the Grant-in-Aid for Scientific Research of the Japan Society for the Promotion of Science (22684004, 23224004).


\begin{deluxetable}{lccccc}
\tabletypesize{\tiny}
\tablecaption{X-ray observations of SN~2005ip}
\tablewidth{0pt}
\tablehead{
\colhead{Date (UT)}&\colhead{Day$^a$}&\colhead{Instrument}&\colhead{Exposure (ks)}&\colhead{Count rate (0.5--10\,keV)}&\colhead{Hardness ratio (2.5--10\,keV)/(0.5--2.5\,keV)}}
\startdata
2007 Feb 14.08 & 474.92& Swift XRT & 8.7 & 4.4$\pm$0.7$\times$10$^{-3}$ & 7.3$\pm$4.3 \\
2008 Jun 6.46  & 798.09& Chandra ACIS-S & 4.7 & 22.4$\pm$0.2$\times$10$^{-3}$ & 2.9$\pm$0.7 \\
2008 Nov 16.03 & 1115.87& Swift XRT & 10.0 & 10.5$\pm$1.0$\times$10$^{-3}$ & 1.9$\pm$0.4 \\
2009 Nov 18.69 & 1483.53& Swift XRT & 6.0 & 10.8$\pm$1.3$\times$10$^{-3}$ & 1.1$\pm$0.3 \\
2012 Mar 20.02 & 2335.86& Swift XRT & 2.4 & & \\
2012 Mar 22.75 & 2338.59& Swift XRT & 0.7 & & \\
2012 Mar 23.28 & 2339.12& Swift XRT & 0.7 & 7.5$\pm$0.9$\times$10$^{-3}$ & 0.5$\pm$0.1 \\
2012 Mar 25.62 & 2341.46& Swift XRT & 3.3 & & \\
2012 Mar 31.57 & 2347.41& Swift XRT & 1.4 & & \\
\enddata
\tablecomments{$^a$Days after explosion.  For {\it Swift} and {\it Chandra} data, the PIs are S.\ Immler and D.\ Pooley, respectively.}
\label{tab:obs}
\end{deluxetable}

\begin{deluxetable}{lccccc}
\tabletypesize{\tiny}
\tablecaption{Spectral properties from the one-component model}
\tablewidth{0pt}
\tablehead{
\colhead{Parameter}&\colhead{2007-02}&\colhead{2008-06}&\colhead{2008-11}&\colhead{2009-11}&\colhead{2012-03}}
\startdata
$N_{\mathrm H}$ (10$^{22}$ cm$^{-2}$)\dotfill 1 & 5.38$^{+3.02}_{-2.18}$ & 3.12$^{+1.44}_{-0.87}$ & 1.35$^{+0.89}_{-0.51}$ & 0.86$^{+1.04}_{-0.68}$ & 0.21$(<0.5)$ \\
\dotfill 2 & 5.35$^{+3.09}_{-2.16}$ & 3.11$^{+1.36}_{-0.87}$ & 1.33$^{+0.88}_{-0.49}$ & 0.85$^{+1.05}_{-0.66}$ & 0.21$(<0.41)$ \\
\dotfill 3 & 5.33$^{+2.10}_{-1.55}$ & 3.08$^{+0.69}_{-0.61}$ & 1.33$^{+0.45}_{-0.38}$ & 0.86$^{+0.55}_{-0.47}$ & 0.22$(<0.46)$ \\
\dotfill 4 & 1.62$^{+0.25}_{-0.22}$ & \multicolumn{4}{c}{Linked to 2007-02}\\
\hline
$kT$ (keV)\dotfill 1 & 68.6$(>6.7)$ & \multicolumn{4}{c}{Linked to 2007-02}\\
\dotfill 2 & 81.5$(>8.5)$ & \multicolumn{4}{c}{Linked to 2007-02 with assumed evolution of $t^{-0.2}$ }\\
\dotfill 3 & 172.6$(>109.6)$ & \multicolumn{4}{c}{Linked to 2007-02 with assumed evolution of $t^{-0.53}$}\\
\dotfill 4 & 135.0$(>52.6)$ & 100.0$(>26.3)$ & 87.8$(>26.3)$ & 7.1$^{+39.7}_{-3.2}$ & 3.0$^{+1.5}_{-0.9}$  \\
\hline
Observed flux ($10^{-13}$\,erg\,s$^{-1}$\,cm$^{-2}$)\dotfill 1 & 5.69$^{+1.67}_{-1.40}$& 6.00$^{+1.03}_{-0.93}$& 8.11$^{+1.42}_{-1.27}$& 7.80$^{+1.79}_{-1.56}$ & 4.46$^{+1.00}_{-0.87}$\\
\dotfill 2 & 5.73$^{+1.68}_{-1.41}$& 6.0$^{+1.0}_{-0.9}$& 8.17$^{+1.43}_{-1.28}$& 7.84$^{+1.80}_{-1.56}$ & 4.48$^{+1.0}_{-0.87}$\\
\dotfill 3 & 5.73$^{+1.68}_{-1.41}$ & 6.04$^{+1.04}_{-0.93}$& 8.17$^{+1.43}_{-1.28}$& 7.80$^{+1.79}_{-1.56}$ & 4.40$^{+0.98}_{-0.86}$\\
\dotfill 4 & 3.91$^{+1.14}_{-0.96}$ & 4.95$^{+0.85}_{-0.77}$& 8.55$^{+1.50}_{-1.34}$& 6.91$^{+1.59}_{-1.38}$ & 3.46$^{+0.79}_{-0.69}$\\
\hline
Absorption-corrected flux ($10^{-13}$\,erg\,s$^{-1}$\,cm$^{-2}$)\dotfill 1 & 9.65$^{+2.83}_{-2.37}$& 8.94$^{+1.54}_{-1.38}$& 10.86$^{+1.90}_{-1.71}$& 9.91$^{+2.28}_{-1.98}$ & 5.03$^{+1.12}_{-0.98}$\\
\dotfill 2 & 9.61$^{+2.82}_{-2.36}$& 8.93$^{+1.54}_{-1.38}$& 10.85$^{+1.90}_{-1.71}$& 9.91$^{+2.28}_{-1.97}$ & 5.04$^{+1.12}_{-0.98}$\\
\dotfill 3 & 9.61$^{+2.82}_{-2.36}$ & 8.93$^{+1.54}_{-1.38}$& 10.85$^{+1.90}_{-1.70}$& 9.91$^{+2.28}_{-1.98}$ & 5.01$^{+1.12}_{-0.98}$\\
\dotfill 4 & 5.34$^{+1.55}_{-1.31}$ & 6.63$^{+1.14}_{-1.02}$& 11.65$^{+2.04}_{-1.83}$& 11.20$^{+2.57}_{-2.23}$ & 7.47$^{+1.71}_{-1.49}$\\
\hline
$L^{a}$ ($10^{41}$\,erg\,s$^{-1}$)\dotfill 1 & 1.43$^{+0.42}_{-0.35}$ & 1.32$^{+0.23}_{-0.20}$ & 1.61$^{+0.28}_{-0.25}$ & 1.47$^{+0.34}_{-0.29}$ & 0.74$^{+0.17}_{-0.15}$  \\
\dotfill 2 & 1.42$^{+0.42}_{-0.35}$ & 1.32$^{+0.23}_{-0.20}$ & 1.60$^{+0.28}_{-0.25}$ & 1.47$^{+0.34}_{-0.29}$ & 0.75$^{+0.17}_{-0.15}$  \\
\dotfill 3 & 1.42$^{+0.42}_{-0.35}$ & 1.32$^{+0.23}_{-0.20}$ & 1.60$^{+0.28}_{-0.25}$ & 1.47$^{+0.34}_{-0.29}$ & 0.74$^{+0.17}_{-0.14}$  \\
\dotfill 4 & 0.79$^{+0.23}_{-0.19}$ & 0.98$^{+0.17}_{-0.15}$& 1.72$^{+0.30}_{-0.27}$& 1.66$^{+0.38}_{-0.33}$ & 1.10$^{+0.25}_{-0.22}$\\
\hline
$C$/d.o.f. \dotfill 1 & \multicolumn{5}{c}{56.1/62}\\
\dotfill 2 & \multicolumn{5}{c}{56.1/62}\\
\dotfill 3 & \multicolumn{5}{c}{56.0/62}\\
\dotfill 4 & \multicolumn{5}{c}{93.1/62}\\
\enddata
\tablecomments{$^a$Calculated in 0.2--10\,keV at a distance of 35\,Mpc.}
\label{tab:param1}
\end{deluxetable}

\begin{deluxetable}{lccccc}
\tabletypesize{\tiny}
\tablecaption{Spectral properties from the two-component model}
\tablewidth{0pt}
\tablehead{
\colhead{Parameter}&\colhead{2007-02}&\colhead{2008-06}&\colhead{2008-11}&\colhead{2009-11}&\colhead{2012-03}}
\startdata
$N_{\mathrm H}$ (10$^{22}$ cm$^{-2}$)\dotfill 1& 7.24$^{+3.27}_{-2.19}$ & 5.06$^{+1.29}_{-1.64}$ & 3.14$^{+1.04}_{-1.34}$ & 2.78$^{+1.15}_{-1.05}$ & 2.32$^{+0.91}_{-1.29}$ \\
\dotfill 2 & 2.94$^{+0.79}_{-0.52}$ & \multicolumn{4}{c}{Linked to 2007-02} \\
\dotfill 3 & 3.26$^{+0.98}_{-0.90}$ &  \multicolumn{4}{c}{Linked to 2007-02} \\
\hline
$kT_l$ (keV)\dotfill 1& 0.22$^{+0.70}_{-0.07}$& \multicolumn{4}{c}{Linked to 2007-02} \\
\dotfill 2 & 0.22$^{+0.07}_{-0.07}$ & \multicolumn{4}{c}{Linked to 2007-02} \\
\dotfill 3 & 0.30$^{+0.15}_{-0.12}$ & \multicolumn{4}{c}{Linked to 2007-02 with assumed evolution of $t^{-0.2}$} \\
\hline
$kT_h$ (keV)\dotfill 1 & 9.19$(>4.63)$ & \multicolumn{4}{c}{Linked to 2007-02}  \\
\dotfill 2 & 88$(>36)$ & \multicolumn{4}{c}{Linked to 2007-02}  \\
\dotfill 3 & 102.3$(>9.0)$ & \multicolumn{4}{c}{Linked to 2007-02 with assumed evolution of $t^{-0.2}$}  \\
\hline
$L_l~^a$ ($10^{41}$\,erg\,s$^{-1}$)\dotfill 1 & 35.0$^{+10.4}_{-9.3}$ & \multicolumn{4}{c}{Linked to 2007-02} \\
\dotfill 2& 0($<$5.2) & 0($<$5.8) & 35.5$^{+20.7}_{-16.6}$& 49.3$^{+34.7}_{-27.0}$& 74.9$^{+26.9}_{-22.6}$\\
\dotfill 3 & 0($<$1.46) & 0($<$2.63) & 19.6$^{+11.3}_{-9.2}$& 35.5$^{+24.1}_{-19.0}$& 82.3$^{+29.7}_{-24.9}$\\
\hline
$L_h~^a$ ($10^{41}$\,erg\,s$^{-1}$)\dotfill 1 & 1.73$^{+0.52}_{-0.43}$ & 1.58$^{+0.29}_{-0.26}$ & 1.90$^{+0.38}_{-0.33}$ & 1.81$^{+0.48}_{-0.41}$ & 0.91$^{+0.26}_{-0.22}$  \\
\dotfill 2 & 1.04$^{+0.30}_{-0.25}$ & 1.29$^{+0.22}_{-0.20}$ & 1.84$^{+0.38}_{-0.34}$ & 1.80$^{+0.52}_{-0.45}$ & 0.80$^{+0.29}_{-0.23}$  \\
\dotfill 3 & 1.04$^{+0.30}_{-0.25}$ & 1.29$^{+0.22}_{-0.20}$ & 1.82$^{+0.38}_{-0.34}$ & 1.76$^{+0.53}_{-0.45}$ & 0.80$^{+0.29}_{-0.24}$  \\
\hline
$C$/d.o.f.\dotfill 1 & \multicolumn{5}{c}{51.6/60}\\
\dotfill 2 & \multicolumn{5}{c}{58.4/60}\\
\dotfill 3 & \multicolumn{5}{c}{58.0/60}\\
\enddata
\tablecomments{$^a$Calculated in 0.2--10\,keV at a distance of 35\,Mpc.}
\label{tab:param2}
\end{deluxetable}

\end{document}